\documentclass[apjl]{emulateapj}

\shorttitle{ACS OBSERVATIONS OF EDIG IN EDGE-ON SPIRALS}
\shortauthors{ROSSA ET AL.}

\newcommand{\lta}{\lesssim}

\begin{document}

\title{{\it HST} ACS OBSERVATIONS OF STAR FORMATION DRIVEN 
OUTFLOWS IN NEARBY EDGE-ON SPIRAL GALAXIES: DEPENDENCE OF HALO MORPHOLOGY 
ON STAR FORMATION ACTIVITY\altaffilmark{1}}


\author{J\"{o}rn Rossa\altaffilmark{2,3},
Michael Dahlem\altaffilmark{4},
Ralf-J\"urgen Dettmar\altaffilmark{5}, and 
Roeland P. van der Marel\altaffilmark{2}}

\altaffiltext{1}{Based on observations made with the NASA/ESA Hubble
Space Telescope, obtained at the Space Telescope Science Institute,
which is operated by the Association of Universities for Research in
Astronomy, Inc., under NASA contract NAS 5-26555. These observations
are associated with proposal 10416.}

\altaffiltext{2}{Space Telescope Science Institute, 3700 San Martin
Drive, Baltimore, MD 21218; marel@stsci.edu}

\altaffiltext{3}{Present address: Department of Astronomy, University
of Florida, 211 Bryant Space Science Center, P.O.Box 112055,
Gainesville, FL 32611-2055; jrossa@astro.ufl.edu}

\altaffiltext{4}{CSIRO/ATNF - Paul Wild Observatory, Locked Bag 194, Narrabri 
NSW 2390, Australia; Michael.Dahlem@csiro.au}

\altaffiltext{5}{Astronomisches Institut, Ruhr-Universit\"at Bochum, 
Universit\"atsstrasse 150/NA7, D-44780 Bochum, Germany; dettmar@astro.rub.de}


\begin{abstract}

We present new high spatial resolution narrowband imaging observations of 
extraplanar diffuse ionized gas (eDIG) in four late-type, actively star 
forming edge-on spirals, obtained with ACS on-board {\it HST}. Our F658N 
(H$\alpha$) observations reveal a multitude of structures on both small and 
large scales. Whereas all four galaxies have been studied with ground-based 
telescopes before, here the small scale structure of the extended emission 
line gas is presented for the first time at a spatial resolution of 0\farcs05, 
corresponding to 5.0\,pc at the mean distance to the target galaxies. The eDIG 
morphology is {\em very} different for all four targets, probably as a result 
of their different levels of star formation activity. We find that the morphology 
of the eDIG, in particular the break-up of diffuse emission into filaments in galaxy 
halos, shows a strong dependence on the level of star formation activity per unit 
area, and eDIG can be arranged into a morphological sequence. The eDIG is 
resolved into an intricate morphology of individual filaments for NGC\,4700 and 
NGC\,7090. The ionized gas in the latter galaxy, being the one with the lowest SF activity 
per unit area in our sample, reveals besides some filaments many prominent bubbles 
and superbubbles. NGC\,4634 and NGC\,5775 have instead a more diffuse layer of eDIG.
The latter two galaxies have the highest SF rate per unit area in our sample and 
the observed morphology suggests that the break-up of the smooth eDIG layer into 
individual resolved filaments occurs only above a certain threshold of SF activity 
per unit area. We note that this is not a distance effect, as NGC\,4634 and 
NGC\,4700 are roughly at the same distance, yet reveal very different eDIG 
morphology. Combined with ground-based data for samples that span a larger range 
of galaxy mass our results indicate that the gravitational potential also plays an 
important role in the eDIG morphology. In low-mass galaxies the gas can be expelled 
due to shallow gravitational potentials more easily and couple with strong 
star formation driven outflows on a local scale. This is in contrast to the more 
massive galaxies, which show smooth eDIG layers, unless they are powered by a 
superwind, as in the case of nucleated starburst galaxies. The simplistic chimney 
scenario is not confirmed from our observations, and future models need to take into 
account the detailed physics of feedback, star formation and magnetic fields. 

\end{abstract}



\clearpage


\section{INTRODUCTION}
\label{s:intro}

During the last decade considerable progress in our understanding of the 
diffuse ionized gas (DIG) in halos of late-type spiral galaxies has been 
achieved. The occurrence of DIG outside of classical \hbox{H\,{\sc ii}} 
regions is firmly believed to be correlated with the star formation activity 
in the galaxy disk. 

The fractional contribution of DIG to the total H$\alpha$ emission in 
galaxies reaches typically values of $\approx$ 25-50\% \citep[e.g.,\,][]{
wal94,fer96,leh95}, and DIG is thus an important constituent of the 
interstellar medium (ISM). Recent studies with a larger sample of galaxies 
from the SINGG survey revealed even slightly higher fractions of $0.59\pm0.19$ 
\citep{oey07}. 

In the theoretical picture the gas, most likely driven by collective 
supernovae, is expelled into the halos of the galaxies. \citet{nor89} 
developed a theoretical model in which the hot gas is transported from the 
star forming regions in the disk via {\em chimneys} into the halo, provided 
the superbubbles created from SNe fulfill the breakout criterion. Depending 
on the strength of the gravitational potential the gas may fall back onto the 
galactic disk, which is described as the {\em galactic fountain} scenario 
\citep{sha76,bre80,avi00}. Starburst driven mass loss, at least for dwarf 
galaxies with shallower potential wells, can be an effective process, as the 
gas can escape more easily in these systems \citep{mac99}.

Several ground-based narrowband imaging investigations have been undertaken 
during the past fifteen years with a boost of investigations occuring in the 
last decade, to study the morphology of DIG in halos of non-starburst 
{\em quiescent} spiral galaxies \citep{det90,ran90,ran92,pil94,ran96,
hoo99,col00,ros00,ros03b,mil03}. Not all of the studied galaxies revealed 
extraplanar DIG (eDIG), but it was shown that extraplanar DIG is a relatively 
common feature among actively star forming galaxies, and about 41\,\% of a 
survey of 74 galaxies have eDIG detections. The most active ones were 
found to possess a bright gaseous halo \citep{ros03a}. Correlations of eDIG 
with other constituents of the ISM exist, such as radio-continuum halos 
\citep[e.g.,][]{dah94,tul00,dah06}, the hot ionized medium (i.e. halos), 
traced by X-ray observations \citep{str04,tul06}, and dust filaments in the 
disk-halo interface \citep{how97,how99,how00,ros03b,ros04,tho04}.   
Starburst galaxies and low luminosity Seyfert galaxies have been investigated as 
well with regard to the diffuse emission \citep{leh95} and \citep{rob07}, 
respectively. 

In general, the spiral galaxies with very high SFR per unit area show a pervasive 
DIG layer (i.e. a halo) with a typical vertical extent of $|z| \sim$ 
2\,kpc. Individual filaments, occasionally reaching distances of up to 6\,kpc, 
are superimposed in a few cases as well \citep{ros03b}. As there is a variety 
of morphological features generally discernible even with low spatial 
resolution ground-based studies (e.g.,\,filaments, plumes, bubbles, shells), 
clearly the need for a more magnified view onto the individual SF regions and 
DIG filaments in edge-on spirals is needed to study the small scale 
structure and to identify the connection points between the filaments and 
their origin within the SF regions in the midplane. Only very few previous 
high spatial resolution investigations were targeted to address these 
questions \citep[e.g.,][]{cec01,cec02,ros04}. For a more thorough overview on 
the DIG topic in general and on phenomena related to these star formation 
driven gaseous outflows we refer to the older, but still quite relevant, 
review articles by \citet{det92,dah97} and also to the more recent review on 
galactic winds by \citet{vei05}.

In this study we present high spatial resolution narrowband H$\alpha$ images 
of four actively star forming galaxies, observed with {\it HST}/ACS to 
unravel the morphology of the gaseous outflows on both small and large scales as 
a function of star formation activity, These data are augmented by broad-band 
{\it HST}/ACS images, whose primary purpose was to serve for the continuum 
subtraction. However, they are also used to study the distribution of dust 
filaments in the disk-halo interface.  

The paper is structured in the following way. In \S~\ref{s:sample} we 
describe how the current {\it HST}/ACS sample was selected, and in 
\S~\ref{s:obs} we present the observations and observing strategy. In 
\S~\ref{s:data} we describe briefly the data reduction techniques and 
analysis of the images. In \S~\ref{s:results} the results are presented, and 
in \S~\ref{s:discuss} we discuss the results. In \S~\ref{s:sum} we finally 
summarize our basic results and state our conclusions. 

\section{THE EDGE-ON GALAXY SAMPLE}
\label{s:sample}
   
\begin{figure}
\includegraphics[angle=270,scale=0.7,clip=t]{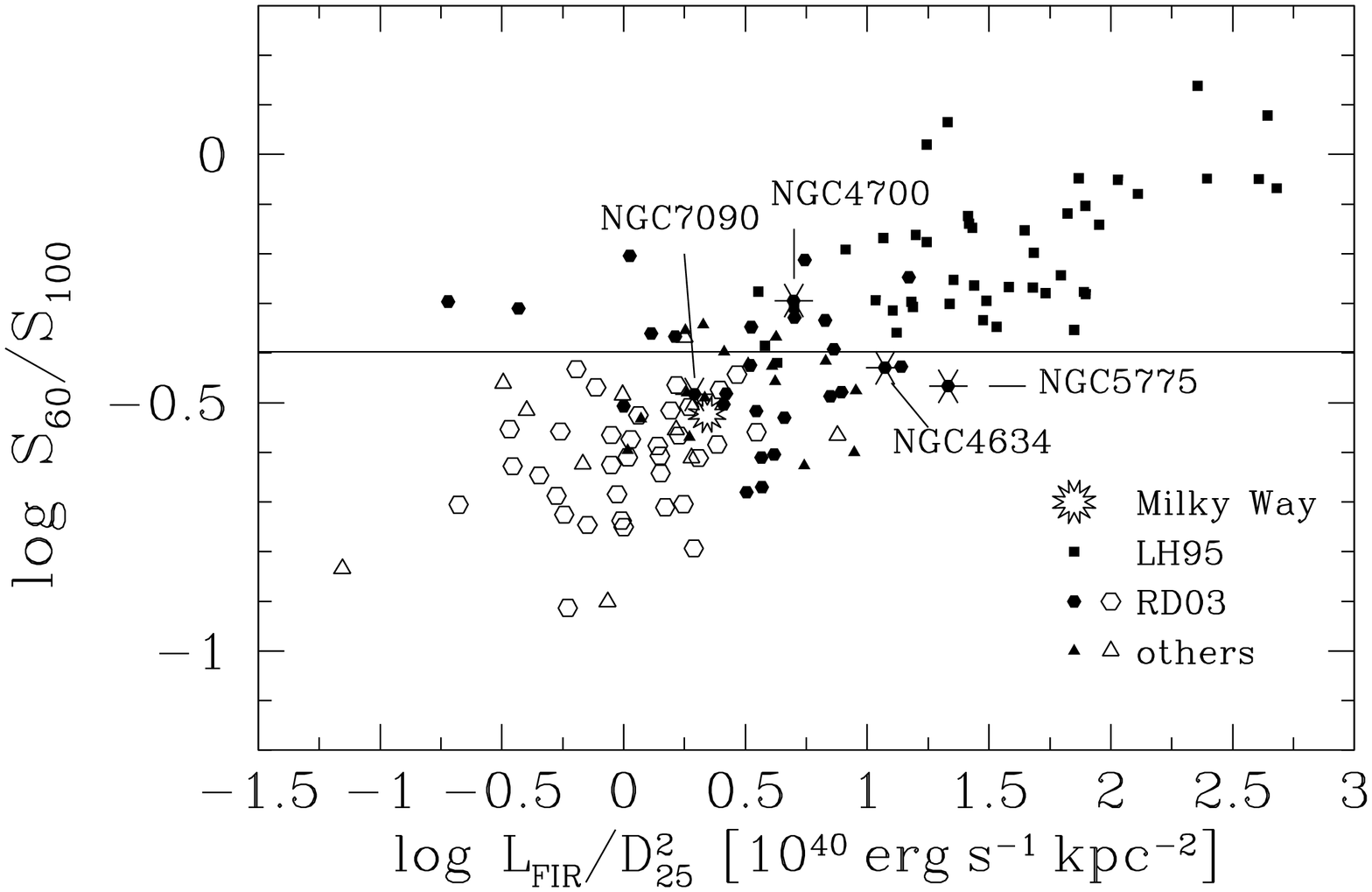}
\caption{Diagnostic DIG diagram \citep[updated and expanded from][]{ros03a} 
with the four galaxies labeled. The horizontal line marks the division of 
IRAS warm galaxies ($S_{60}/S_{100} \geq 0.4$). Filled symbols denote 
galaxies with eDIG detections, whereas open symbols denote galaxies without 
eDIG detections. Note that the Milky Way was labeled with an open symbol for 
better visibility in this diagram, even though it reveals extraplanar 
emission.\label{f:dddhst}}
\end{figure}

The small edge-on galaxy sample presented here is a sub-sample of four 
galaxies from our previous H$\alpha$ survey \citep{ros00,ros03a,ros03b}. 
The basic properties of these galaxies are listed in Table~\ref{t:obs}. 
The galaxies were selected to be observed with {\it HST}/ACS, because they 
have very different levels of energy input into their disk ISM (expressed
as $L_{\rm{FIR}}/D^2_{25}$ and listed with other properties in Table~\ref{t:prop}). 
All four targets have eDIG detections \citep{ros00,ros03a,ros03b}, but ground-based 
imagery suggests different halo morphologies that can be further studied at the highest 
possible resolution only by means of {\it HST}/ACS H$\alpha$ imagery. Galaxies 
with different energy input rates were chosen to study a possible dependence 
of the halo morphology on the level of star formation in the underlying 
disks. Naively, one might expect galaxies with energy input levels near 
the threshold for the creation of halos (see Figure~\ref{f:dddhst}) to 
exhibit only individual, possibly filamentary extraplanar structures.
On the other hand, galaxies with widespread, high-level star formation 
are expected to have a multitude of filaments, if they are associated 
with individual sites of star formation in the underlying disks.

We avoided galaxies with prominent dust lanes and those at low Galactic 
latitudes so as to have low foreground opacities for studies of the 
optical emission line gas.

We note here that two galaxies in the present sub-sample, NGC\,4634
and NGC\,5775, have close interaction partners (within two Holmberg
radii), while the remaining two, NGC\,4700 and NGC\,7090, don't.

\section{OBSERVATIONS AND OBSERVING STRATEGY}
\label{s:obs}

The observations of the four actively star forming spirals were carried 
out between June 23 and August 21 2005 with the Advanced Camera for Surveys 
(ACS) onboard {\it HST} (GO\#10416; PI: J.~Rossa). All observations were carried out 
using the Wide Field Channel (WFC), which yields a $202\arcsec \times 202\arcsec$ 
field of view. This is achieved by employing two SITe 2048$\times$4096 pixel CCDs 
with a pixel scale of 0\farcs05 per pixel. For more details we refer to the ACS 
instrument handbook \citep{gon05}. The narrowband observations using the 
F658N filter were acquired in three {\it HST} orbits for each galaxy, and the 
broadband images (to be used for continuum subtraction and to study the dust 
morphology in the galaxies) were obtained with the F625W filter in one {\it 
HST} orbit each. In total, 16 {\it HST} orbits were allocated to obtain deep 
continuum subtracted H$\alpha$ images of the four target galaxies. The 
observational details, including observation dates, integration times, and 
filter characteristics, are listed in Table~\ref{t:exp}. 

The individual exposures were taken in dither mode, to allow for good 
cosmic ray removal and to fill in the CCD gaps. We have used a different 
dither pattern for the broadband and narrowband observations. For the 
narrowband (F658N) observations we have used a 3-point dither pattern with 
a 4-point sub-pattern, spread over three {\it HST} orbits for each galaxy. 
The broadband (F625W) observations were obtained in a 2-point dither with 
a 2-point sub-pattern. The first pattern in each configuration applies 
a jump over the CCD gaps, and the secondary pattern is applied to allow 
for good cosmic ray and hot pixel removal.     

\section{DATA REDUCTION AND ANALYSIS}
\label{s:data}

\subsection{Data reduction}
\label{ss:datared}

The {\it HST}/ACS data were calibrated using the {\it CALACS} and 
{\it Multidrizzle} packages. The former removes the various instrumental 
signatures, including bias correction and flat-fielding. The latter is 
used to correct for the geometric distortion of the ACS camera and also 
performs cosmic ray rejection and finally combines the images, obtained in 
the specific dither pattern. For further details we refer to the ACS data 
handbook \citep{pav05} and to \citet{koe05} for a description of multidrizzle. 

There are a few detector artifacts visible at very low intensities. Among them 
are some faint stripes along CCD columns visible across the images. These are 
charge transfer trails of hot pixels, cosmic rays, and other sources. However, 
these CTE losses are well defined features, which are located at a constant position, 
and whose pattern is repeated with a constant distance and position angle. On very 
faint levels, when viewing the entire frame in low resolution on a computer screen, 
effects of cross talk are discovered. They have their origin in the four CCD 
quadrants corresponding to the four amplifiers of the detector array. The features 
though look somewhat different from the usual cross talk features seen in 
\citet{gia04}. The features in our images are not inverted in intensity and 
thus can easily be mistaken for faint diffuse emission. However, the distance 
and position angle of this feature in the continuum subtracted H$\alpha$ image of 
NGC\,4634 mimics the shape of the galaxy at a very faint level, and its endpoint 
coincides with the one end of the galaxy, at a position that has an offset and 
inverted position. As an almost identical feature was detected in the image of 
NGC\,5775 as well, which looked very reminiscent, we conclude, that this is 
indeed related to the detector, and not a real feature of the diffuse 
emission. At higher contrast the structure resolves and is barely visible at 
all.    
 
\subsection{Calibration and Analysis of the Broadband Imagery}
\label{ss:broad}

The broadband (F625W) images of our target galaxies were obtained for two 
purposes. The primary aim was to subtract the continuum from the on-band 
(F658N) images, which contain some continuum flux. Secondly, they allow a 
detailed comparison of extraplanar dust features with the DIG morphology, 
which are best seen in unsharp-masked versions of the broadband images. 
These two physically distinct ISM phases \citep[see\,][]{how00} can be 
studied in detail to search for possible correlations between the two 
extraplanar ISM constituents on small scales. To enhance the contrast between 
the dust structures and the background light from the galaxy more easily, we 
have constructed unsharp-masked versions of our broadband F625W (SDSS $r$) 
images. These were obtained by dividing the F625W images by a smoothed 
version of the same images. The latter ones were created by using a 
Gaussian filtering technique with a kernel size adopted to account for 
the difference in distance among the sample galaxies.   

\subsection{Calibration and Analysis of the Narrowband Imagery}
\label{ss:narrow}

In order to calibrate the narrowband images the broadband and narrowband 
images were first aligned by measuring the positions of stars in the field of 
view. After alignment the shifted broadband images were checked by blinking 
them on a computer screen against the narrowband images. In order to obtain a 
continuum-free H$\alpha$ image, we scaled and subtracted each combined 
broadband F625W image from the respective narrowband F658N image. This was 
done in a similar fashion as it was applied in our previous studies 
\citep[e.g.,][]{ros00,ros04}. The use of individual foreground stars instead 
proved difficult, as there were only limited suitable stars in the given 
aperture visible. More importantly, they also showed too much color variation 
so that the scaling factor could not be determined very accurately. Therefore, 
we chose selected reference regions within the galaxy that were not affected 
by emission. For this purpose we closely compared the F625W and F658N images 
to determine suitable regions within the disk, and also relied on previous 
ground-based images.  

Several of the continuum subtracted images show black-white point source pairs
which are due to slight imperfections in alignment and differences in the PSF. 
However, they do not affect our discussion and understanding of the extended 
emission. Due to the given FWHM of the used narrowband filter, it should be noted 
that emission from the adjacent [\ion{N}{2}] doublet is included in the F658N 
filter passband. For simplicity, however, we refer for the remainder of the 
paper to an H$\alpha$ image, whenever we speak of a continuum subtracted 
F658N filter image.

\section{RESULTS}
\label{s:results}

In this section we describe in detail the morphology of the four edge-on 
spirals as revealed from both the broadband and narrowband images. 
Specifically, we describe the DIG morphology visible in the emission line 
images. We also comment on the dust features visible in the broadband 
images and on a newly detected dwarf irregular galaxy in the halo of NGC\,4634.

\begin{figure}
\includegraphics[angle=0,scale=1.0,clip=t]{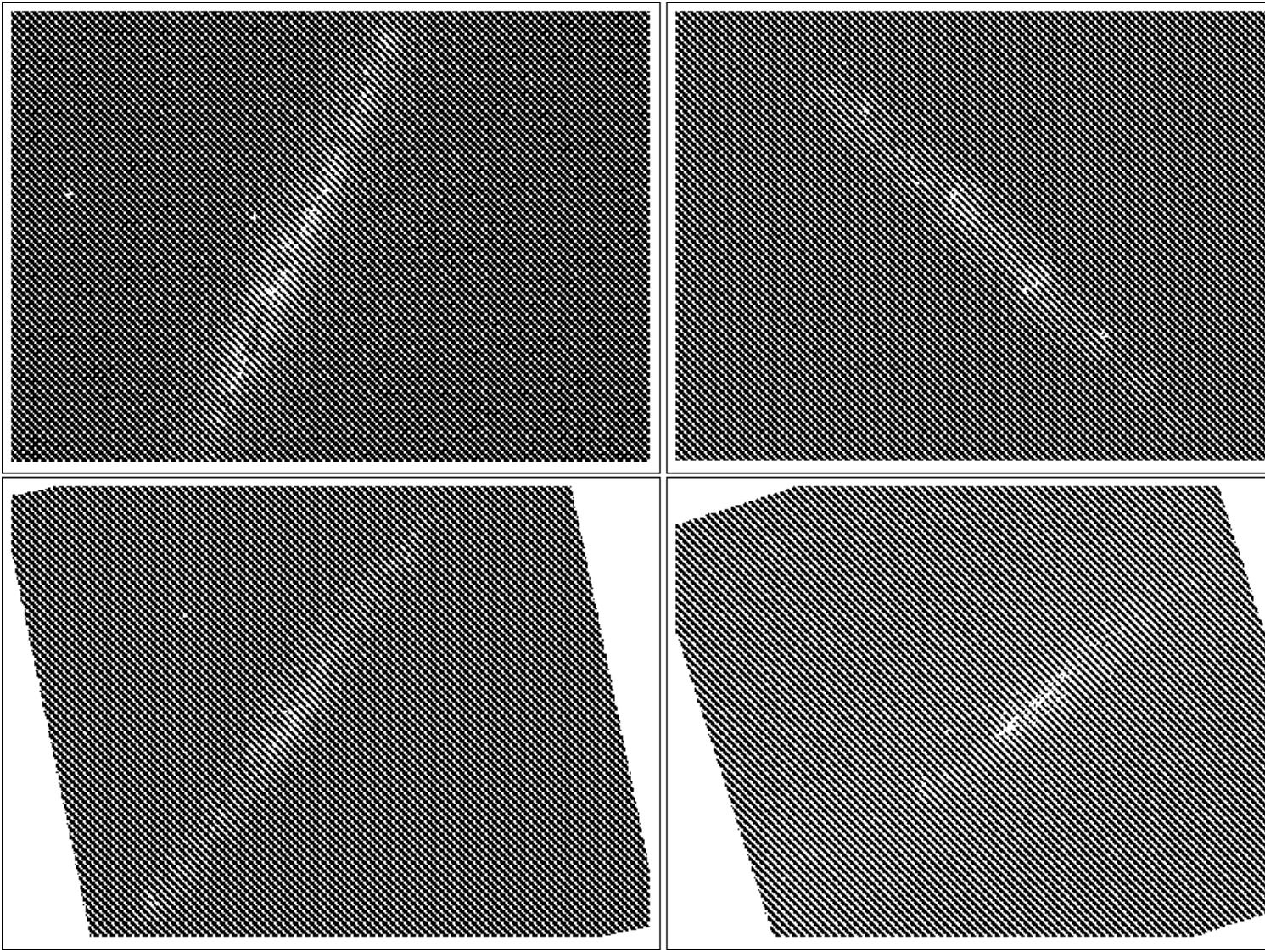}
\caption{Color composite images of NGC\,4634 (top left), NGC\,4700 (top right), 
NGC\,5775 (bottom left) and NGC\,7090 (bottom right), as observed with {\it HST}/ACS. 
The spatial resolution is $0\farcs05$\,pix$^{-1}$. The red color represents the 
continuum subtracted H$\alpha$ image, the green color represents the H$\alpha$ image, 
both of which trace the emission line gas, whereas the blue color represents the continuum 
image (F625W = SDSS $r$), which indicates the stellar component. The orientation of the 
individual figures is the usual, with north to the top and east to the left. 
\label{f:colorfigures}}
\end{figure}

The layout of the images is as follows. In \S~\ref{ss:broadimage} we present the 
three-color images (Figure~\ref{f:colorfigures}), which show both the stellar and the gas 
distribution for all of the galaxies. In \S~\ref{ss:narrowimage} we present the emission 
line images (Figures~\ref{f:n4634sixpanel}, \ref{f:n4700sixpanel}, \ref{f:n5775sixpanel} 
and \ref{f:n7090sixpanel}). Each of the four figures consists of six images. The overview 
images are shown in the upper left corner, and the individual enlargements in the remaining 
five subpanels. For NGC\,4700 we show an enlargement of the three-color image 
(Figure~\ref{f:n4700acs}), revealing the central region. The unsharp-masked images (tracing 
the dust distribution), which are discussed in \S~\ref{ss:dust}, are shown in a separate 
figure (Figure~\ref{f:dust4}), as well as the dIrr galaxy near NGC\,4634 
(Figure~\ref{f:dIrr}), presented in \S~\ref{ss:dIrr}.

\subsection{ACS broadband imagery}
\label{ss:broadimage}

We have combined the broadband and narrowband images to better display the 
interplay of gas and stars in the disk-halo region. Our three-color 
images (F625W = blue; F658N = green; continuum subtracted F658N = red) of 
NGC\,4634, NGC\,4700, NGC\,5775 and NGC\,7090 are presented in Figure~\ref{f:colorfigures}. 
They provide us with an overview of the disk morphology, in particular the stellar 
distributions (i.e. star densities) within each galaxy. This information can later be used 
to investigate potential dependencies of halo properties on the {\it local} level of disk 
activity.

\paragraph{NGC\,4634}
Star forming regions (and regions of increased stellar densities) are quite 
wide-spread throughout the disk of NGC\,4634. The broadband image shows a 
number of \ion{H}{2} regions shining through the galaxy's dust lane. Around 
the high surface-brightness inner disk fainter ``emission'' is found, probably 
from stars in outer spiral arms that may be slightly warped. No significant 
stellar overdensities are found in these outer regions.

\paragraph{NGC\,4700}
The stellar disk of NGC\,4700 looks more irregular than that of
NGC\,4634. The dust morphology is patchy, rather than forming a
continuous band along the disk plane. The most prominent star concentrations 
are found in the central area. The northeastern peak of the stellar 
distribution has a higher stellar density, than the southwestern peak. 
Both regions seem to be more or less equidistant from the center of the 
galaxy. There is one particular region located within the disk that has quite 
prominent extinction by dust. It is also interesting to note that the patchy 
dust lane is slightly skewed with respect to the stellar disk plane. 

\paragraph{NGC\,5775}
NGC\,5775 has a well-defined disk with evidence of widespread, 
high-level star formation, as indicated by copious regions of 
increased stellar density. Its dust lane is similar to that of
NGC\,4634 and more pronounced than that of NGC\,4700. The dust lane is 
most pronounced in the outer disk regions (i.e. at large galactocentric 
radii). However, NGC\,5775 has a relatively compact nuclear region, that 
has the bulk of the stellar densities. 

\paragraph{NGC\,7090}
NGC\,7090's disk is similar to that of NGC\,4700, with a slightly more 
prominent dust lane and an apparently wider, more amorphous overall 
appearance. The dust lane is very filamentary. Most parts of the dust lane 
are located in the regions above the midplane. Not much dust is present in the 
southern part of the disk (below the midplane). The distribution of its stellar 
component is asymmetric. The maximum is located to the southeast of the center, but 
there are several peaks of stellar density distributions along the disk.

\subsection{ACS emission line images}
\label{ss:narrowimage}

\begin{figure}
\includegraphics[angle=0,scale=1.0,clip=t]{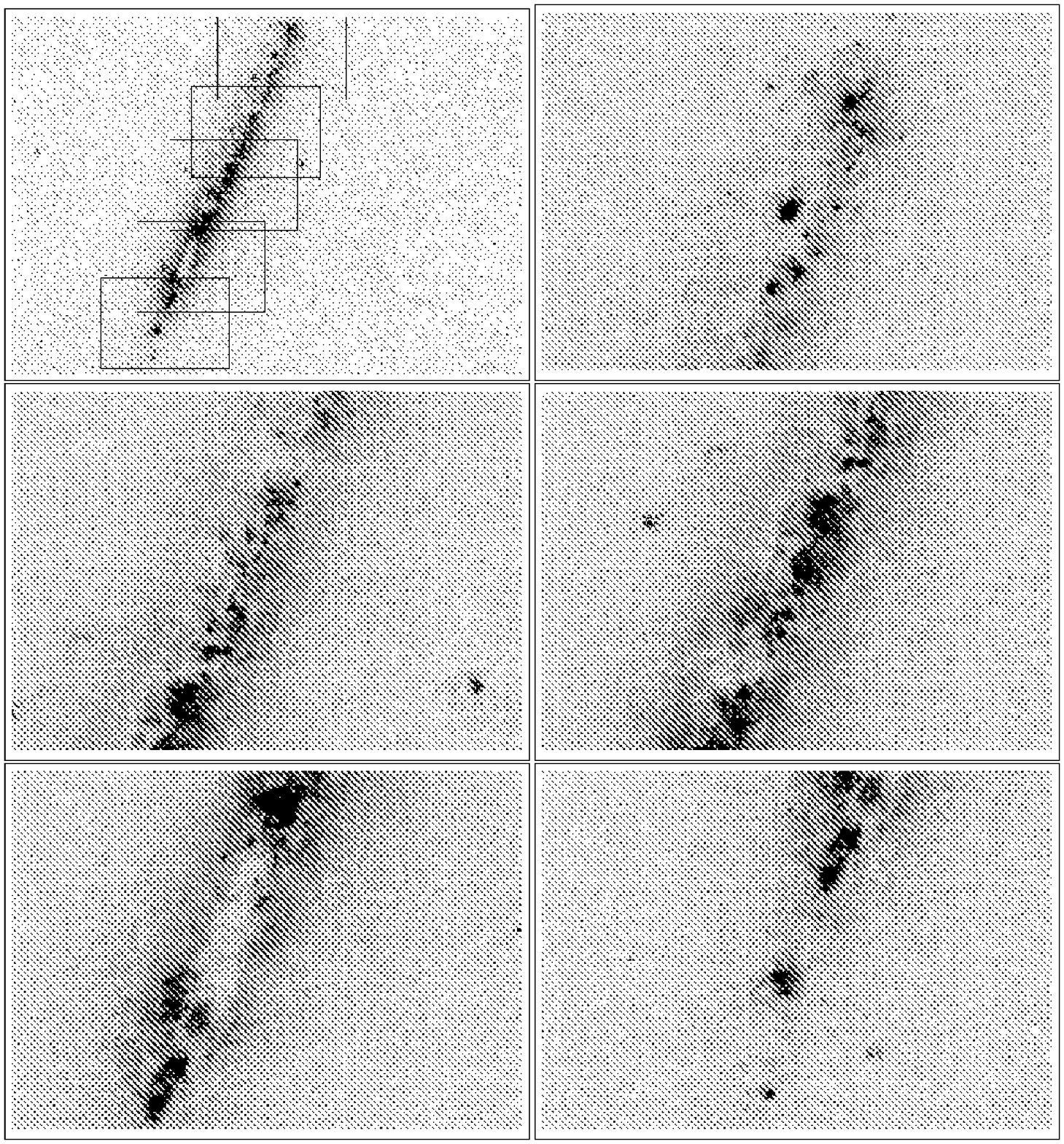}
\caption{{\it HST}/ACS continuum subtracted H$\alpha$ image of NGC\,4634. 
The orientation is with N to the top and E to the left. The labeled boxes in the 
overview image (top left) refer to the individual regions A, B, C, D and E, which are 
shown as enlargements in the subsequent panels to reveal a more detailed view. The boxes 
overdrawn on the image to the top left refer to the image areas shown in the following 
subpanels: A = top right, B = middle left, C = middle right, D = bottom left, and E = bottom 
right. \label{f:n4634sixpanel}}
\end{figure}

\begin{figure}
\includegraphics[angle=0,scale=1.0,clip=t]{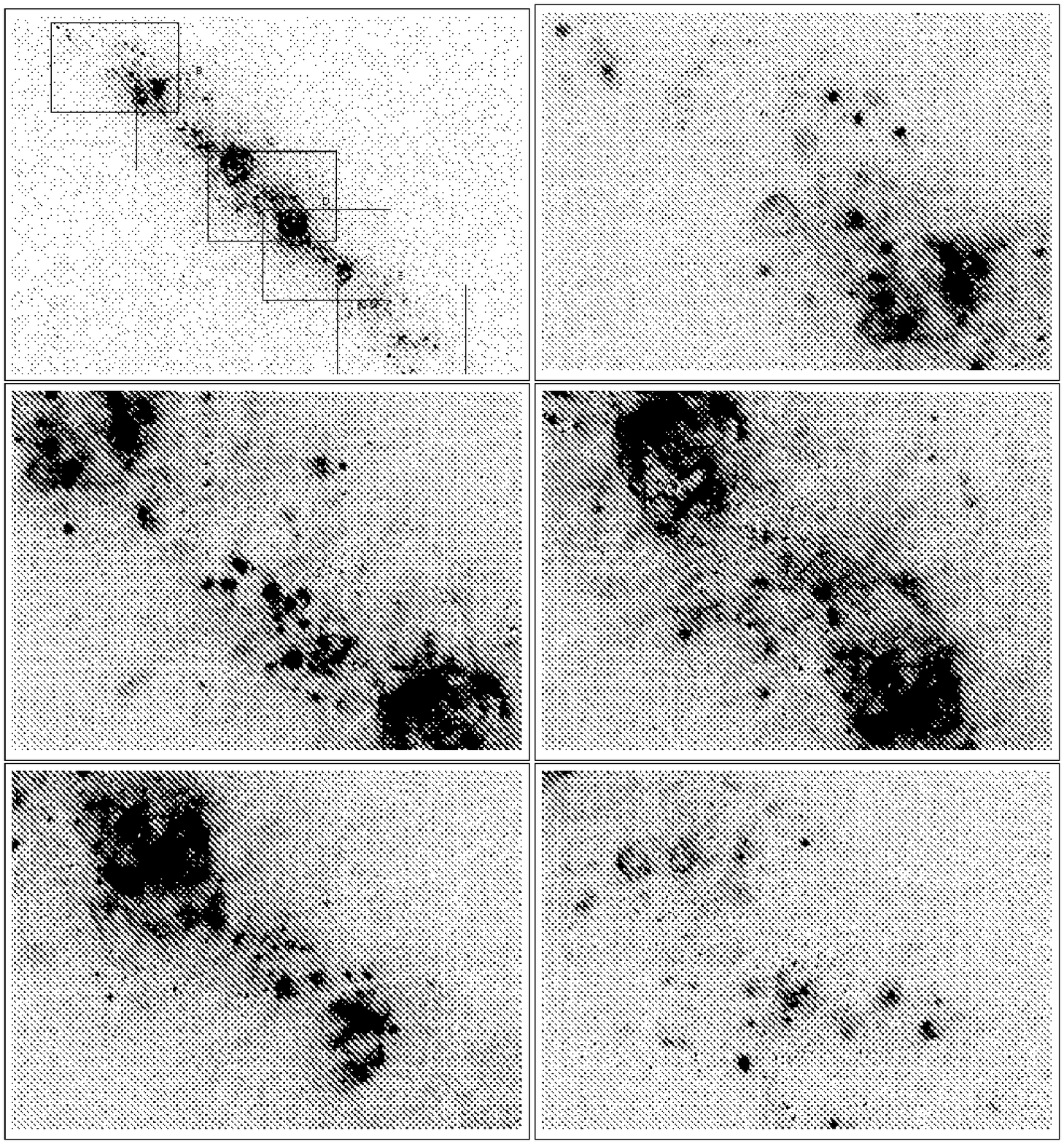}
\caption{{\it HST}/ACS continuum subtracted H$\alpha$ image of NGC\,4700. 
The orientation is with N to the top and E to the left. The labeled boxes in the 
overview image (top left) refer to the individual regions A, B, C, D and E, which are 
shown as enlargements in the subsequent subpanels to reveal a more detailed view. The boxes 
overdrawn on the image to the top left refer to the image areas shown in the following 
subpanels: A = top right, B = middle left, C = middle right, D = bottom left, and E = bottom 
right. \label{f:n4700sixpanel}}
\end{figure}

\begin{figure}
\includegraphics[angle=270,scale=0.8,clip=t]{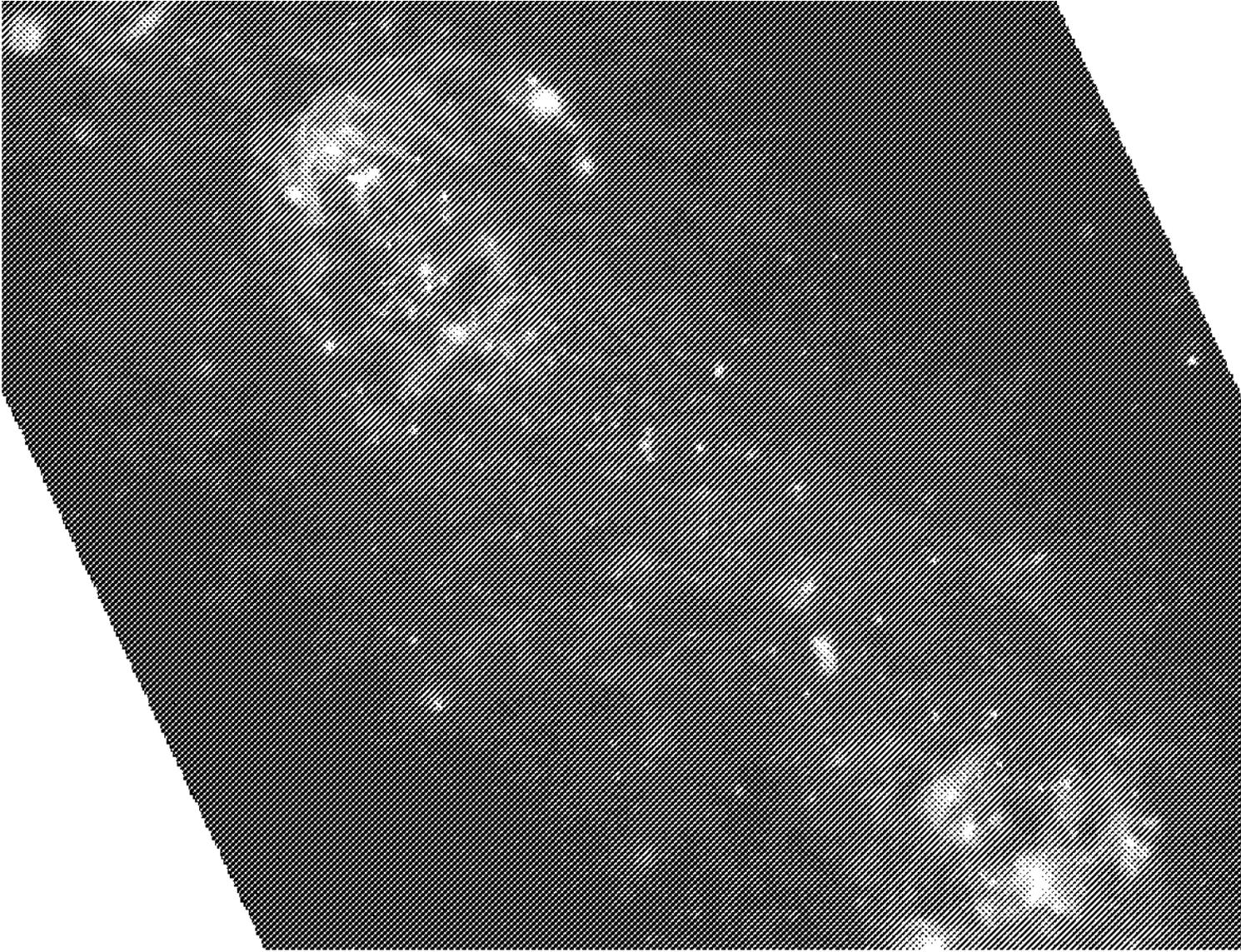}
\caption{Color composite image of the central region in NGC\,4700, as observed 
with {\it HST}/ACS. A very complex DIG morphology is revealed, with many 
individual superimposed filaments protruding from the disk region far into 
the halo. The spatial resolution is $0\farcs05$\,pix$^{-1}$. The red color 
reflects the continuum subtracted H$\alpha$ image, the green color reflects 
the H$\alpha$ image, both of which trace the emission line gas, whereas the 
blue color represents the continuum image (F625W = SDSS $r$), which indicates the 
stellar component. The orientation is the usual way, with north to the top and east 
to the left.\label{f:n4700acs}}
\end{figure}

\begin{figure}
\includegraphics[angle=0,scale=1.0,clip=t]{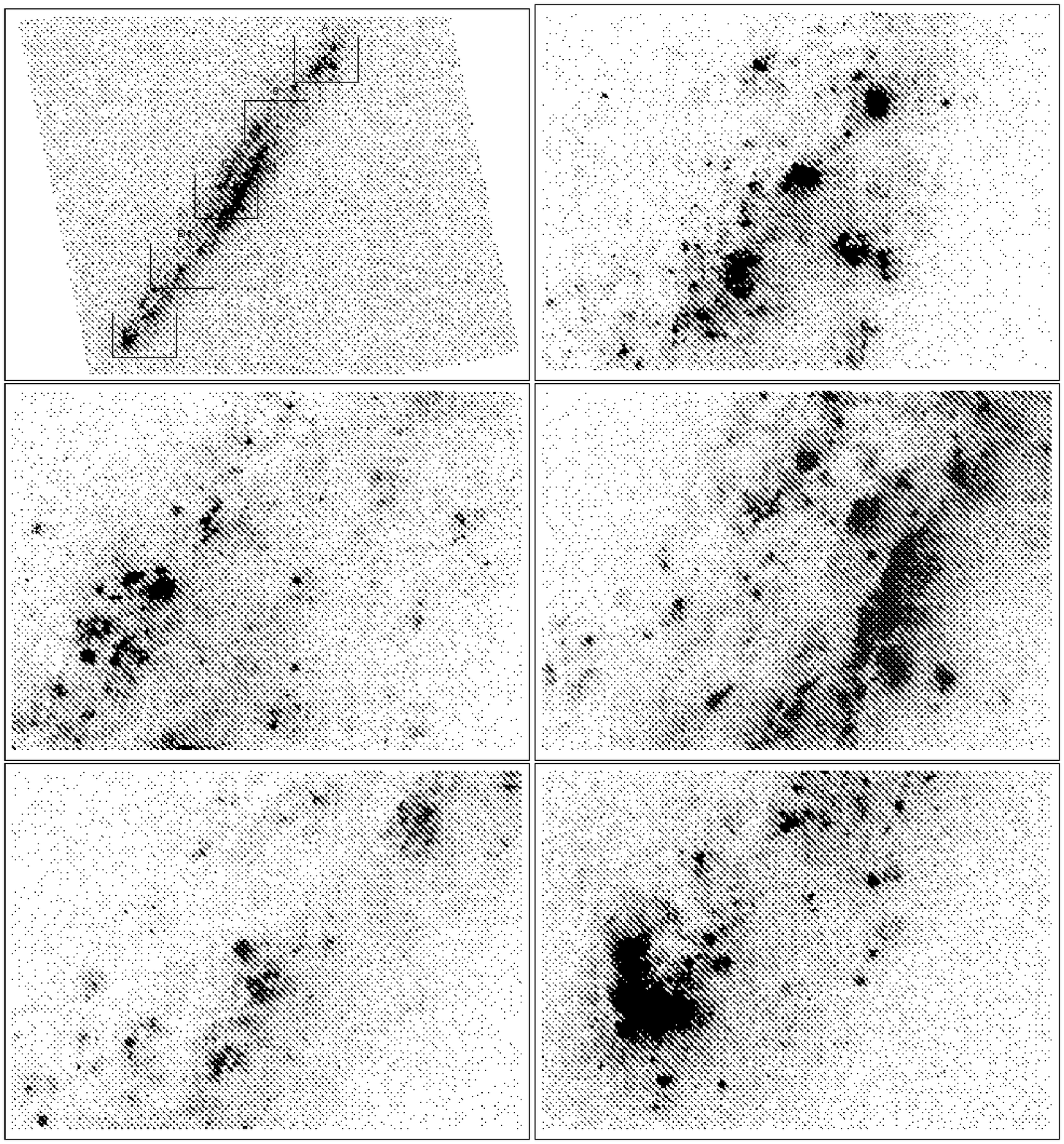}
\caption{{\it HST}/ACS continuum subtracted H$\alpha$ image of NGC\,5775. 
The orientation is with N to the top and E to the left. The labeled boxes in the 
overview image (top left) refer to the individual regions A, B, C, D and E, which are 
shown as enlargements in the subsequent panels to reveal a more detailed view. The boxes 
overdrawn on the image to the top left refer to the image areas shown in the following 
subpanels: A = top right, B = middle left, C = middle right, D = bottom left, and E = bottom 
right.\label{f:n5775sixpanel}}
\end{figure}

\begin{figure}
\includegraphics[angle=0,scale=1.0,clip=t]{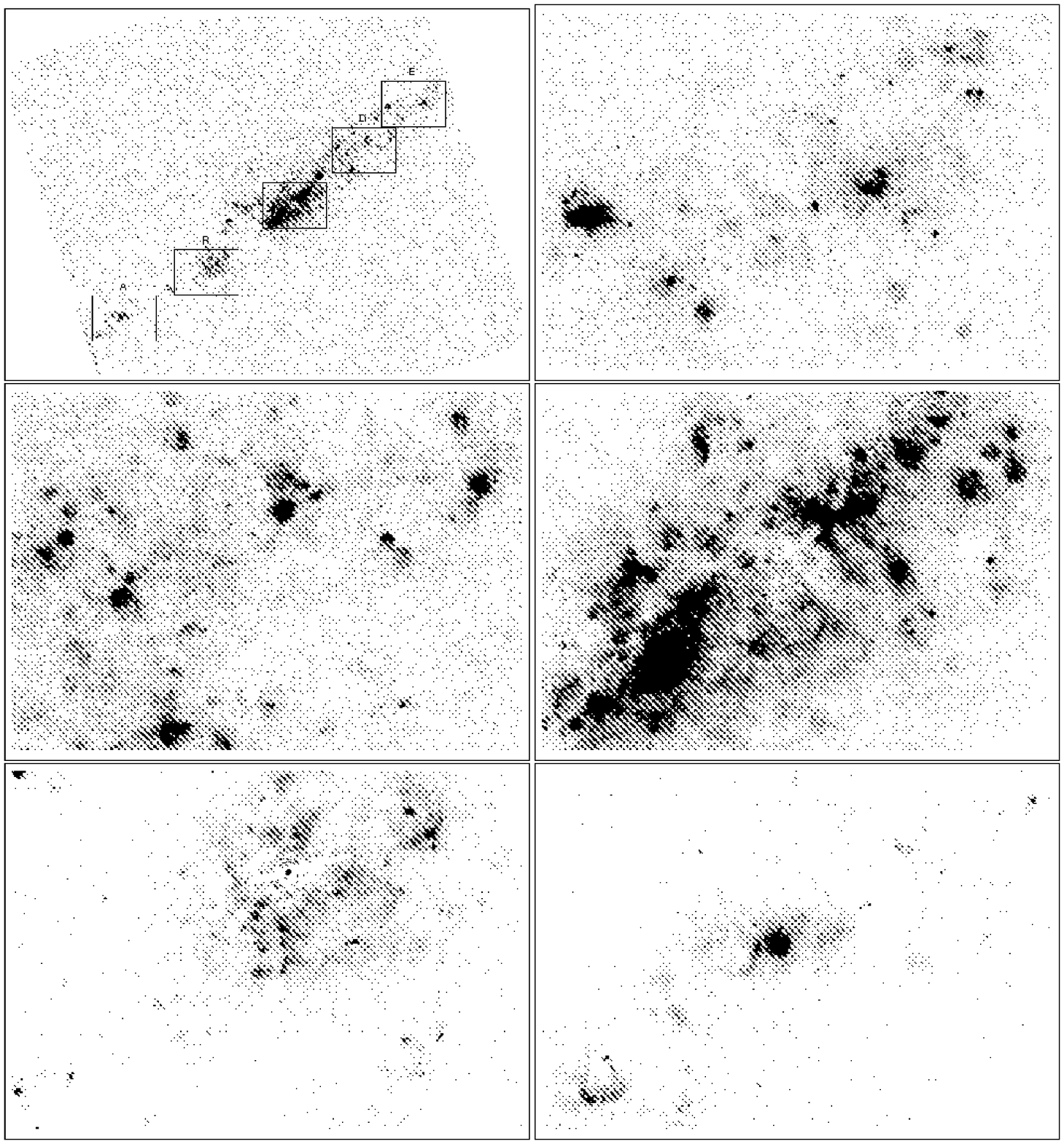}
\caption{{\it HST}/ACS continuum subtracted H$\alpha$ image of NGC\,7090. 
The orientation is with N to the top and E to the left. The labeled boxes in the 
overview image (top left) refer to the individual regions A, B, C, D and E, which are 
shown as enlargements in the subsequent panels to reveal a more detailed view. The boxes 
overdrawn on the image to the top left refer to the image areas shown in the following 
subpanels: E = top right, D = middle left, C = middle right, B = bottom left, and A = bottom 
right. \label{f:n7090sixpanel}}
\end{figure}

\begin{figure}
\includegraphics[angle=0,scale=1.0,clip=t]{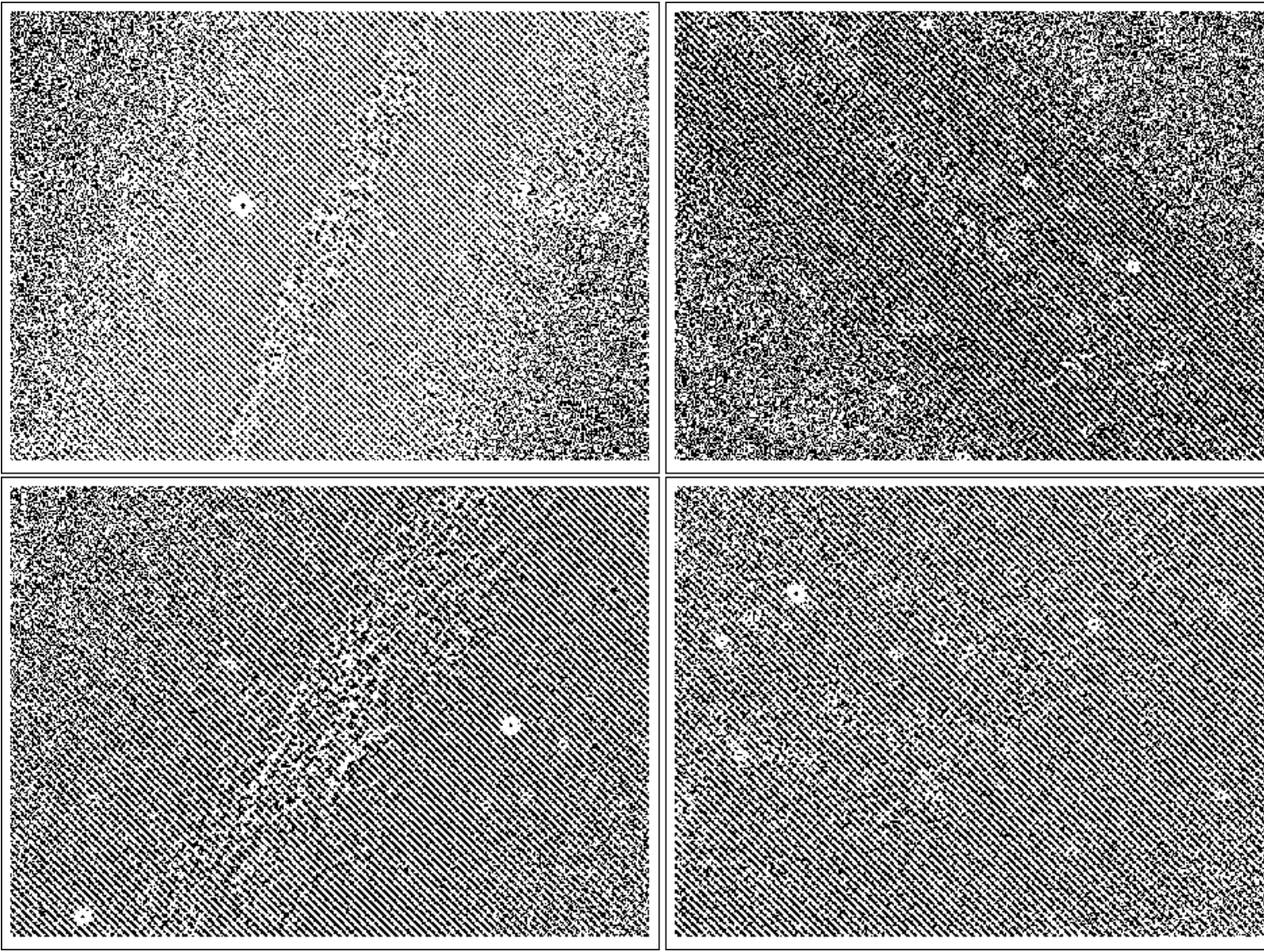}
\caption{Unsharp-masked images revealing the extraplanar dust distribution 
in NGC\,4634 (top left), NGC\,4700 (top right), NGC\,5775 (bottom left) and 
NGC\,7090 (bottom right). The global orientation of the dust filaments in 
NGC\,4634 is mostly perpendicular to the disk on both sides of the disk at 
high $|z|$. At lower $|z|$ the orientation of the dust filaments is random. 
Due to the lack of a strong dust lane in NGC\,4700 (and the general lack of 
patchy dust in this galaxy), the unsharp-masked image does not reveal 
much insight into the dust distibution. Also, the presence of many stars 
that are clearly resolved within NGC\,4700, complicates the interpretation 
of the distribution of extended dust filaments. Only the central part of 
NGC\,5775 is shown here for better visibility. There is a filamentary web of 
dusty structures revealed. The global orientation of the dust filaments are mostly 
parallel to the disk on the one side of the disk, whereas the other side has mostly 
an orientation perpendicular to the disk. Generally, though, at lower $|z|$ the orientation 
is random. For NGC\,7090 also only the central part is shown here. Due to the lack 
of a pronounced dust lane, the unsharp-masked image does not reveal much unsight 
into the extraplanar dust distribution. \label{f:dust4}}.
\end{figure}

\begin{figure}
\includegraphics[angle=0,scale=0.95,clip=t]{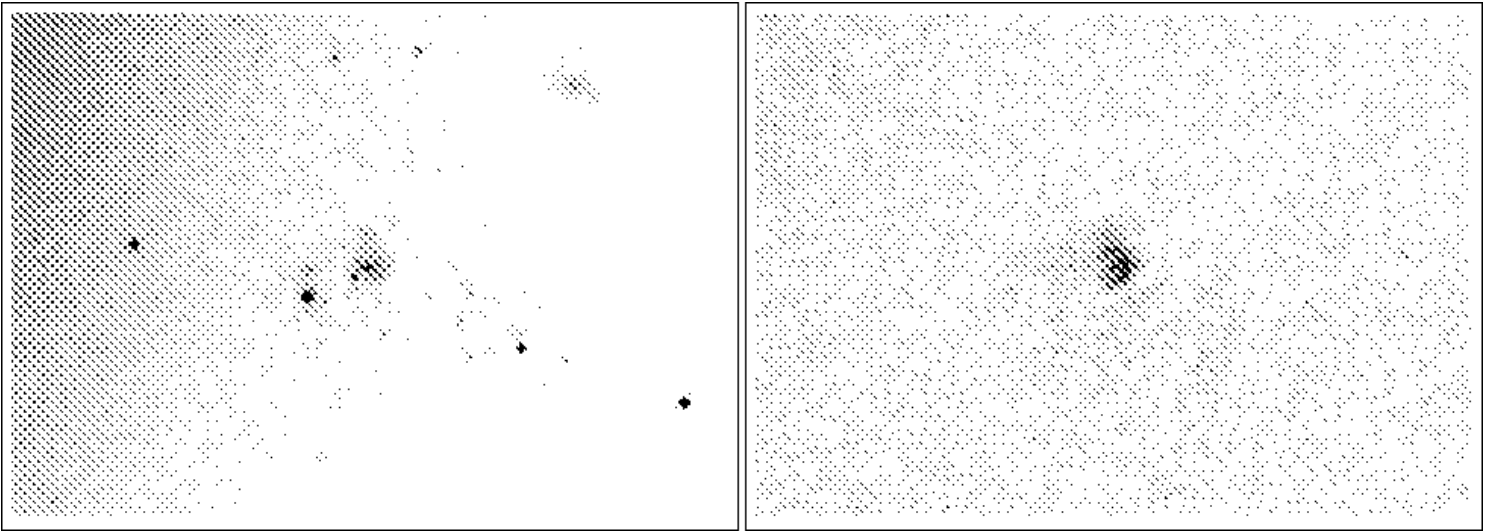}
\caption{The tidally disrupted dIrr galaxy, $\rm{J}\,124239.58$+141751.83, 
in the halo of NGC\,4634, as revealed by {\it HST}/ACS. The left panel shows 
the broadband image (F625W), the right panel shows the continuum subtracted 
H$\alpha$ image. The field of view measures $17\farcs8 \times 12\farcs3$, which 
corresponds to 1650\,pc$\times$1140\,pc at the distance to NGC\,4634. The orientation 
is the usual with north to the top and east to the left. The diagonal color gradient 
in the broadband image is not a flatfield artifact. It reflects the increase of galaxy 
light from the halo region in the left side of the figure in this color stretch. 
\label{f:dIrr}}
\end{figure}

As mentioned above, these images, called ``H$\alpha$'' images, 
include unknown contributions from the adjacent [\ion{N}{2}] lines and 
thus trace the warm interstellar emission line gas in general. Both the 
disk and the extraplanar gas can be studied in all four target 
galaxies. Images showing the full extent in the ACS field-of-view are shown for
each galaxy in the upper left panel of Figures~\ref{f:n4634sixpanel}, 
\ref{f:n4700sixpanel}, \ref{f:n5775sixpanel} and \ref{f:n7090sixpanel}. 

In addition, close-up presentations of individual areas in the four
target galaxies are presented below in the five remaining sub-panels of each 
figure. We show five individual fields for each galaxy. Comments on the observed 
morphology are summarized in Table~\ref{t:morph}. We also include some information 
on previous ground-based studies and related disk-halo investigations from other 
wavelength regimes that have a direct connection to our present study. 

\paragraph{NGC\,4634}

This Virgo cluster member edge-on spiral was previously imaged in the DIG 
context in H$\alpha$ using ground-based data \citep{ros00}. A diffuse halo 
was detected with an extent of 1.2\,kpc above/below the galactic plane. 

Extraplanar emission was also confirmed spectroscopically \citep{tul00a,ott02}. 
In addition to revealing eDIG, extraplanar dust was also detected at high 
galactic latitudes of about 1\,kpc \citep{how99,ros03b}. 

As already discussed in \citet{ros03a}, the presence of an H$\alpha$ halo 
is almost always accompanied by a hot gaseous (X-ray) halo, as high SF rates 
and high $S_{60}/S_{100}$ are strong indicators of halo emission. Not 
surprisingly, a hot tenuous and far more extended X-ray gaseous halo was found 
in NGC\,4634 in the soft energy band, making use of XMM-Newton observations 
\citep{tul06}. A unique feature among the studied sample of nine edge-on 
spirals is the absence of hard X-ray emission in NGC\,4634. 

Our ACS H$\alpha$ image of NGC\,4634 (see Figure~\ref{f:n4634sixpanel}) shows widely 
distributed star formation in its disk, as already suggested by the broadband 
image introduced above. Individual \ion{H}{2} regions are seen through gaps in 
the dust lane. But there are no prominent H$\alpha$ bubbles visible.

In the halo extended H$\alpha$ emission is found that is, despite
the superb spatial resolution of {\it HST}/ACS, unresolved (see 
Figure~\ref{f:n4634sixpanel}). No filaments, loops or similar structures 
are discernible, apart from a few minor exceptions at large galactocentric radii. 

\paragraph{NGC\,4700}

This southern edge-on spiral was previously imaged by means of ground-based 
H$\alpha$ observations \citep{ros03b}. A bright halo with superimposed filaments 
reaching into the halo out to $|z| \approx 2-3$\,kpc was detected. The filaments 
appear to emanate from the brightest SF regions, anchored within the disk. 

Also, a radio-continuum halo was detected \citep{dah01}, revealing similar 
shapes and maxima at the optical SF counterparts. The extent of the 
cosmic-ray halo is much larger than the optical halo. 

Spectra of the galaxy itself have been published by \citet{kew01}, which 
show besides prominent H$\alpha$ and [\ion{N}{2}] emission relatively 
strong [\ion{O}{3}] emission lines. However, \citet{kew01} do not assign 
a classification for this galaxy, which they list as IRAS\,$12465-1108$. 
An imaging HST survey of \citet{mal98} lists NGC\,4700 as an \ion{H}{2} galaxy.
Spectral classifications in the NASA Extragalactic Database (NED) range 
from \ion{H}{2} region to Sy\,2 type, so there is not a clear-cut classification 
available for this galaxy as of yet.

NGC\,4700 (Figure~\ref{f:n4700sixpanel}) displays an enormous wealth of H$\alpha$ 
emitters in its disk plane with a large number of bright star forming regions 
and many bubbles interspersed. Contrary to the rather irregular appearance
of its stellar disk in the broadband image (cf. \S~\ref{ss:broadimage}), 
the H$\alpha$ emission distribution looks much more confined to what 
looks like a thin disk. The ACS H$\alpha$ image (specifically the central 
region displayed in Figure~\ref{f:n4700acs}) is by far the most spectacular one 
in our sample. Given the distance of NGC\,4700, which is very similar to 
NGC\,5775, this represents no bias in terms of resolution (i.e. proximity). 
The H$\alpha$ image reveals a very complex morphology. There are numerous faint 
filaments protruding into the halo region. 

The circumnuclear region does not host strong SF activity, and it seems as if 
the bulk of the emission has been ejected into two lobe regions. These are, 
despite the higher densities, located within the galactic plane (roughly 
aligned with the disk plane) and not perpendicular to it. Nonetheless, considerable 
structure (e.g.,\,filaments) is also revealed perpendicular to the disk north 
and south of the nuclear region. Currently we do not have any kinematic 
information of the lobe regions and the galaxy in general, so we can only 
speculate about the possible origin of the H$\alpha$ lobes. But it seems most 
likely as if these were created by a powerful mass and energy ejection from the 
nuclear region, which may host an AGN. 

On top of the disk emission an equally astounding wealth of extraplanar
H$\alpha$ emission is found, with a large number of long, thin filaments
emanating vertically from the central disk plane. Several of these filaments
can be traced back to individual knots of H$\alpha$ emission in the central 
part of the disk, suggesting that the gas is energized from within the disk 
plane. The majority of the long filaments are emanating from and around 
regions close to the nucleus. However, we note that from optical observations 
(both R-band and H$\alpha$) no clear indication of a nucleus can be assessed. 

Instead two giant SF complexes (almost equidistant from the center) are 
located within the disk plane, which resemble in appearance the lobe 
regions of a radio jet (on smaller scales, though). However, no jet is visible, 
nor is there any clear indication of a distinct nucleus. These shell-like structures 
are located within the disk, and show a radially expanding structure within the 
disk, where the pressure gradient is much higher as opposed to moving perpendicular 
to it (such as the winds in nucleated starbursts). The very edges of the disk show 
substantial amounts of SF. Many superbubbles and shells are detected, and there does 
not appear to be an abrupt cut-off in star formation within the disk. The expanding 
shells seem indicative of an AGN (Seyfert type), however no clear classification has 
been established. 
 
\paragraph{NGC\,5775}

NGC\,5775 is a well studied edge-on spiral galaxy in the disk-halo context, 
which was imaged extensively by ground-based observations in H$\alpha$ 
\citep{leh95,col00,tul00,ros03b}. Extraplanar emission was imaged out 
to $|z|\lta$\,5\,kpc above the galactic plane. It is the galaxy with the 
highest $L_{\rm{FIR}}/D^2_{25}$ ratio in our current sample and it is a 
starburst-type galaxy, although it is morphologically quite different from 
that of nucleated starbursts. A gaseous halo with individual H$\alpha$ spurs or 
filaments, associated with the X-shaped structure of the large scale magnetic 
field, was discovered \citep{tul00}. 

EDIG has been detected spectroscopically even out to $|z|\sim$\,9\,kpc 
\citep{ran00,tul00}. In addition, extended radio continuum emission has been 
detected as well \citep{hum91,dur98,tul00}. A tenuous X-ray halo, spatially 
coexistent (and more extended) than the radio continuum and H$\alpha$ halo, 
was detected with Chandra \citep{str04} and with XMM-Newton \citep{tul06}. 
NGC\,5775 appears to be interacting with its face-on companion NGC\,5774, which 
was first indicated by \ion{H}{1} measurements \citep{irw94}.

In NGC\,5775 quite a large number of prominent \ion{H}{2} regions is found 
in its disk plane, together with a number of bubbles 
(see Figure~\ref{f:n5775sixpanel}). The disk is surrounded by extended H$\alpha$ 
emission that is not resolved into individual structures. The extraplanar 
H$\alpha$ emission is strongest and widest in the central area, where the level 
of star formation in the disk is also the highest, but appears to be truly 
diffuse. The high spatial resolution of our {\it HST}/ACS images does not allow 
for the detection of the very faint surface emission of the filaments. Unfortunately, 
the background artifacts discussed in \S~\ref{ss:datared} did not allow to bin the data.

\paragraph{NGC\,7090}

This southern edge-on spiral is the galaxy with the smallest 
$L_{\rm{FIR}}/D^2_{25}$ in our current sample, and it was imaged in 
H$\alpha$ before, using ground-based data \citep{ros03b}. These images already 
revealed an inhomogeneous layer of extended DIG, which seems to correlate spatially 
with the star formation regions within the disk. In addition to this H$\alpha$ study, 
an extended radio continuum halo was detected as well \citep{dah01}, based on ATCA 
observations with different configurations at frequencies of 1.43\,GHz and 2.45\,GHz. 
The measured scale-heights for these observations range from 1.19 to 2.16\,kpc. 

The overview image of our ACS H$\alpha$ emission line image of NGC\,7090 (see 
top left panel of Figure~\ref{f:n7090sixpanel}) confirms the asymmetry of the 
star formation activity in its disk very clearly. To the east of the nuclear 
region there is a big apparent gap in the H$\alpha$ emission distribution. In 
general, the eastern half of the galaxy disk is considerably less active than the 
western one. Part of this can be attributed to dust, which is prominently visible 
in the broadband image.

The image also reveals that there are not just \ion{H}{2} regions 
(as traced by diffuse, high surface-brightness H$\alpha$ emission) 
in the disk, but lots of individual, resolved bubbles in addition. 
These are not only present in the more active western half, but also 
to the east, where a few individual bubbles are found out to large 
galactocentric radii.

Images of individual regions in H$\alpha$ reveal the presence of a multitude 
of conspicuous, {\it resolved} individual structures, including partial 
loops and filaments perpendicular to the disk plane. A few of the extended 
filaments are detected in the southern part of the galaxy. These connect 
visibly to the bright star forming regions in the disk (see middle right panel 
of Figure~\ref{f:n7090sixpanel}). 

It is obvious that the H$\alpha$ emission distribution is apparently
widest above regions with the most active star formation in the disk.
There are a multitude of bubbles and supershells detected (see 
top right and middle left panels of Figure~\ref{f:n7090sixpanel}).

\subsection{Dust features in the ACS broadband images}
\label{ss:dust}

Filtering the low spatial frequencies out of the F625W broadband images, 
one can study the smaller-scale absorption features caused by dust in front 
of background stellar emission.

Apart from the gaseous constituents of the ISM in galaxies, dust is also 
present on both small and large scales. To understand the morphology of
the extraplanar ISM in our target galaxies, it is interesting to compare 
the structure of extraplanar dust to that of the gaseous emission filaments. 
For this purpose, similarly to the previously studied cases, we have created 
unsharp-masked versions of the broadband F625W (SDSS $r$) images to enhance 
the contrast between those absorbing high spatial frequency dust structures 
and the smoother distribution of the background light from the galaxy's
stellar populations.  

Despite the fact that our target galaxies were selected to have a less dusty plane 
than galaxies such as NGC\,891 \citep{how97,ros04}, there are still sufficient 
numbers of small-scale dust filaments for such a comparison. Due to the fact 
that we only obtained broadband images in one waveband (the basic reason being 
that these are only used for the continuum subtraction), we cannot quantify 
the physical properties of the dust. We therefore only qualitatively describe the 
extraplanar dust content in these galaxies with respect to the eDIG morphology.  

The images are shown in Figure~\ref{f:dust4}. They reveal that the 
distribution of the dust filaments is also quite different among the four 
target galaxies. NGC\,4634 and NGC\,5775 show the most prominent features, 
whereas NGC\,4700 and NGC\,7090 have the least extraplanar dust features. 
For NGC\,4700, in particular, but also to some extent for NGC\,7090, the presence 
of many resolved stars in the disk make an analysis relatively difficult, as the 
smoothing process creates some artifacts around the stars, which overlap prominently 
in the crowded field. Therefore, a direct one-to-one comparison of dust filaments 
with emission line features (e.g., filaments) is only possible for NGC\,4634 and NGC\,5775.

However, as presented in \S~\ref{ss:narrowimage}, there are a few large-scale 
H$\alpha$ filaments discernible at high galactic latitudes 
in NGC\,4634 and NGC\,5775. The distribution of dust structures in NGC\,4634 
show both alignment parallel and perpendicular to the disk, whereas 
generally at higher $|z|$ the dust (sometimes very filamentary) is aligned 
perpendicular to the disk. But it should be noted that this is only true 
for the western side of the galaxy. The distribution of dust features on 
the other side (east of the center) is somewhat more confined to the disk and 
oriented more parallel than perpendicular, but mostly a random mix of 
orientations. It should also be noted that the dust features are not 
correlated on a one-to-one basis with the H$\alpha$ filaments. This was 
already stated earlier \citep{how99,ros03b}, based on groundbased observations.

\subsection{Dwarf irregular galaxy in the halo of NGC\,4634}
\label{ss:dIrr}

With our high spatial resolution {\it HST}/ACS observations, we are able to 
re-investigate the extraplanar blob, detected by us in the halo of NGC\,4634, 
which was named Patch\,1 \citep{ros00}. We had concluded from our previous 
ground-based observations that this patch may either be an extraplanar SF 
region or a dwarf galaxy in the halo of NGC\,4634, given that the FWHM of the 
used filter excluded the possibility of a foreground or background object. 

Our deep {\it HST}/ACS broadband and narrowband observations now clearly indicate 
that this patch is a galaxy. The narrowband F658N observations reveal a complex system 
composed of several knots associated with gas, reminiscent of a dwarf 
irregular (dIrr) galaxy (see Figure~\ref{f:dIrr}). The broadband image reveals 
several individual star associations that are resolved in this dIrr galaxy, 
and the major morphology looks reminiscent of a satellite merger galaxy, 
showing hints of tidal streams owing to the gravitational attraction from 
NGC\,4634. Given the peculiar morphology, this dIrr galaxy is possibly already 
in the early process of producing a minor merger with its parent galaxy 
(NGC\,4634). It should be noted that the broadband image shows a very thick 
stellar disk of NGC\,4634, unlike any of the other studied edge-on galaxies. 
This is consistent with scenarios in which disk thickening is the result of
a previous interaction, possibly previous minor mergers \citep[e.g.,][]{yoa06,wys06}. 

Augmented longslit spectra (R.-J. Dettmar et al. 2008, in preparation) seem 
to strengthen this interpretation of Patch\,1 being a dIrr galaxy, which is 
deficient in [\ion{N}{2}] and [\ion{S}{2}], based on emission line 
diagnostics. Henceforth, we label this galaxy as $\rm{J}124239.58$+141751.83, 
which refers to the coordinates of the brightest H$\alpha$ emission region 
in its central part. We will report on the physical properties and stellar 
populations of this interesting dIrr galaxy in detail in a future paper.   

\subsection{Ancillary data}

The {\it HST}/ACS data presented above were obtained as part of a
multi-wavelength observing campaign. For the galaxies presented here
we have other, independent measurements of various phases of the ISM
in their halos published by us previously, including X-ray data 
\citep[][Ehle et al.~2008, in preparation]{tul06} and radio-continuum 
data \citep{tul00,dah01}. 
A correlation study of the ACS data with other tracers of extraplanar 
gas in this sample galaxies will be presented separately, as some of the  
observations are still pending.

\section{DISCUSSION}
\label{s:discuss}

\subsection{EDIG morphology as a function of SF activity}
\label{ss:morphvssfprop}

Over roughly the past decade and a half evidence has been mounting that
extended gaseous halos around late-type spiral galaxies are formed as a 
consequence of the star formation activity in the underlying disks. A high 
IRAS $S_{60}/S_{100}$ integral flux ratio $\geq 0.4$, which is a measure 
of a high mean dust temperature in galaxies, was identified as a very reliable 
indicator of massive star formation \citep{hec90}.

In such a scenario, emission features in gaseous halos are expected to be 
closely tied to the energy input rate, which is required to exceed a threshold 
to drive outflows perpendicular to the disk plane in the first place. This 
was first addressed by \citet{dah95} for radio data and later derived from 
optical narrowband imagery \citep{ran96,ros03a}. 
More recently, \citet{dah06} have shown that there is a correlation of the 
existence of galaxy halos (detected in radio continuum and H$\alpha$ 
observations) not only with the global energy injection rate, but also with 
the {\it local} energy input level in a galaxy and, in addition, with its mass 
density. We will investigate in the following whether the eDIG properties of 
the galaxies in the present sample can be explained in this context.

The two galaxies with filamentary eDIG structures are NGC\,4700 and NGC\,7090.
As tabulated in Table~\ref{t:prop}, these are the two galaxies at the lower end 
of the energy input range ($L_{\rm FIR}/D_{25}^2$). On the other hand, NGC\,4634 
and NGC\,5775 display only unresolved extraplanar H$\alpha$ emission. These two 
galaxies have higher energy input rates from massive star formation in their disks. 
Accordingly, one might expect them to have a larger number of individual sites of 
star formation and thus also a larger amount of associated halo emission.

One might naively expect this difference to be an observational artifact,
based on the assumption that physically larger and more massive galaxies
just have more of everything. More star formation regions lead to more
filaments emanating from the disks, potentially leading to a quasi-diffuse
eDIG structure caused by the superposition of many individual filaments.
However, there are no indications of filaments being detected in the more
massive galaxies of the present sample (NGC\,4634 and NGC\,5775) and also
other galaxies studied previously, in areas where the spatial density 
of star-forming regions decreases, e.g., near the radial edges of their disks. 
The difference in eDIG morphologies is equally unlikely to be caused by a
distance bias, because NGC\,4700 (with a filamentary eDIG) and NGC\,5775
(with a diffuse eDIG) are at roughly equal distances (see Table~\ref{t:obs}). 
Again, this makes it unlikely that the diffuse eDIG is the result of an
unresolved superposition of many filaments.

A first indication at what might cause this difference can be gleaned by
studying various properties of the sample galaxies. NGC\,4634 and NGC\,5775 
are the two galaxies with diffuse eDIG, one of them is also the most massive 
galaxy in the present sample (see Table~\ref{t:prop}). From the width of the 
\ion{H}{1} emission line we derive an estimated total mass for NGC\,5775 of about 
$1.91\times10^{11}\,{\rm M}_\odot$. Interestingly though, in contrast, the 
mass of NGC\,4634 is almost a factor of two below that of NGC\,7090. So this 
complicates the interpretation that the mass alone is a sensitive indicator 
for the morphology of eDIG. 
 
The general trend and the earlier work by \citet{dah06} indicate that the mass 
(or mass density) of a galaxy is an important control parameter for the 
morphology of the eDIG. Low mass galaxies such as the dwarf NGC\,55 
\citep[e.g.,][]{tul03}, NGC\,2188 \citep{dom97} and the blue compact 
dwarf galaxy IZw18 \citep[e.g.,][]{hun95} show filamentary structure, whereas 
the more massive galaxies (with widespread star formation, not the powerful 
nucleated starbursts; e.g., NGC\,891 \citep{ros04}) tend to have relatively 
smooth layers of eDIG. An explanation for this could be the fact that the gas 
can overcome the gravitational potential wells of the galaxies more easily. 
The range of masses spanned by our sample is rather small (see Table~\ref{t:prop}), 
so it may not be surprising that any trends with mass are less clear in our sample 
than those observed in samples that contains both dwarf and giant galaxies.

Of our targets only NGC\,7090 and, to a lesser degree, NGC\,4700, contain bubbles, shells 
or supershells. In NGC\,7090 a large number of such features, of various sizes, are found.
Also, the SF regions in this galaxy are loosely scattered within the disk, and the major 
outflows are confined to only one side above the disk (see descriptions of the eDIG 
morphology in the individual regions within each galaxy listed in Table~\ref{t:morph}). 

Among our studied sample, the resolved small-scale DIG halo morphology is showing 
a clear dependence on the star formation rate per unit area $(L_{\rm{FIR}}/D^2_{25})$. 
That is, galaxies with $L_{\rm{FIR}}/D^2_{25} \geq 10$ (in units of 
$\rm{10^{40}\,erg\,s^{-1}\,kpc^{-2}}$), show a smooth, diffuse DIG halo, as opposed 
to the galaxies with values $\leq 6$, which show primarily filamentary DIG. The studied 
galaxies thus can be arranged into a {\it sequence} of eDIG diffuseness as a function 
of $L_{\rm{FIR}}/D^2_{25}$. More generally speaking, there is a morphological DIG sequence 
among SF galaxies observed evolving from galaxies which reveal mostly bubbles and 
supershells to galaxies which show filamentary emission and to cases which have smooth 
eDIG layers (i.e. halos). Where there are bubbles (i.e. enclosed structures) the breakout 
condition cannot have been met yet.

\subsection{The role of the intergalactic environment}
\label{ss:environ}

Although it has been established that eDIG originates from the SF regions 
within the disk, the galactic environment in which the individual galaxies 
reside, is of course of importance, too. It may, e.g., contribute to the 
extent of the emission that is detected. 

Of the four galaxies studied here, two (NGC\,4634 and NGC\,5775) have nearby
companion galaxies (i.e. they form galaxy pairs). NGC\,4700 and NGC\,7090, in 
contrast, are rather isolated spirals. 

Our current sample is not large enough to draw any statistical conclusions.
But since one galaxy from each category has a filamentary and the other a 
diffuse eDIG morphology, the presence of a nearby companion does not appear 
to dominate the eDIG properties in these cases. However, significant 
exceptions may be expected in dense group environments, where galaxies 
experience effects such as ram pressure stripping \citep[e.g.,][]{ken99}.

\subsection{Clues from observations at other wavebands}
\label{ss:multiwave}

It is important to ask whether or not there seems to be a more global connection 
between the various ISM constituents that are responsible for driving a kiloparsec 
outflow or wind. Obviously, there is a connection linking the various ISM phases 
as described for modest galaxy samples \citep[e.g.,][]{tul06,dah06}, but the 
details are far from being well understood. 

From the sample of \citet{tul06}, it was concluded that the presence of a diffuse 
ionized gas halo was always accompanied by a grand scale X-ray halo, which 
generally revealed that the hot tenuous gas is far more extended than the DIG. 
In those cases also a cosmic ray halo, traced by radio-continuum observations, was 
discovered. Linking the cosmic-ray halo to the DIG halo, showed clear evidence 
that the mass surface density was a good proxy for the presence of a gaseous halo. 
The mass of the stellar disk ($M_{\rm K}$), derived from K-band observations, is divided 
by the area of the optical disk ($A_{25} = \pi r_{25}^2$), which yields the mass surface 
density. 

For NGC\,4700 a value of $M_{\rm K}/A_{25} = 17.4\,{\rm M_{\sun}\,pc}^{-2}$ was 
determined \citep{dah06}. Despite having higher values than the more quiescent 
galaxies, this is a rather modest number compared to local starburst galaxies 
which can have mass surface densities of up to eight times as high.

However, does the mass density critically control whether a galaxy has a 
kiloparsec outflow or rather a more filament-driven DIG morphology? Our present 
sample is certainly too small to answer this question unambiguously, as two galaxies 
each fall into the category of either having a filamentary DIG morphology, or showing 
a more smooth DIG layer. 

\subsection{Observations confront Theory: Where have all the Chimneys gone?}
\label{ss:obsvstheo}

One of the greatest challenges any theoretical concept poses is how 
accurately it can reflect or predict the observational details. In terms of 
the chimney model \citep{nor89}, past observations from different wavebands, 
both in galactic and extragalactic studies, have failed to reveal many of 
these structures that were predicted. More specifically, the number of 
predicted chimneys is of the order of several hundred in a galaxy of modest 
SF activity. Yet, only very few chimneys have been detected, if any at all. 
So the prediction falls short by at least two orders of magnitudes. Previous 
reports include a Galactic chimney in the Perseus arm \citep{nor96} and 
the Galactic chimney GSH\,$277+00+36$ \citep{mcg03} as well as evidence for 
a chimney breakout in the Galactic supershell GSH\,$242-03+37$ \citep{mcg06}. 
Extragalactic chimneys or structures identified as such so far have been 
observed only in the starburst galaxy M\,82 \citep{wil99,gar01}. But we have 
to carefully consider whether this is due to an observational bias or attributed 
to certain physical contraints. 

Previously, NGC\,891 was imaged in H$\alpha$ \citep{ros04}, where no evidence 
was found for the chimney mode. However, the situation in NGC\,891 is 
complicated by the fact that it has a very prominent dust layer, which 
possibly could have obscured any chimney structures (i.e. chimney walls) 
in the lower disk-halo interface ($|z| \leq 500$\,pc). Therefore, those 
structures could have gone undetected even if they existed. Our currently 
studied galaxies by and large have a much less dusty galactic midplane, so 
the effects of dust obscuration should be considerably mitigated.  

However, from our imaging studies it is obvious that there are no clear 
indications for the chimney mode in these actively star forming galaxies. 
In NGC\,4634 and NGC\,5775 there are no such structures observed in H$\alpha$. 
Are there indications of other related phenomena? Evidently, there are 
several features (e.g.,\, curved filaments) visible whose morphology show an almost 
closed loop, indicative of the galactic fountain scenario. However, without 
detailed knowledge of the kinematics (e.g.,\,mapping of the velocity field) 
we cannot say what parts of the looped filament are still moving upward and 
what parts possibly are already descending. Furthermore, multiple episodes of SF 
driven outflows may have altered the geometry of the outflows along the line of 
sight. Hence the chimney structures may look like sheets, or picked up more 
complex structures generated by overlapping filaments. Nonetheless, it seems 
obvious that these features are associated with an SF driven outflow. 

Recent models by \citet{tas06} simulated the star formation in galactic disks 
involving feedback by SNe. The density, temperature, and pressure maps (see 
their Figure~6) show copious amounts of warm and hot gas being driven into the 
halo regions, extending out to a few kiloparsecs. In particular, those simulations 
look very reminiscent to the morphology seen in our H$\alpha$ image of 
NGC\,4700 on a global scale. More recent models of a galactic wind as seen in 
starburst galaxies \citep{coo08} replicate very detailed structures including the 
filamentary morphology seen as for instance in M\,82. While considerable progress 
has been achieved in modeling even small scale structures in a magnetized ISM 
\citep[e.g.,][]{avi05}, more refined simulations are necessary in order to 
compare them to the high spatial resolution multi-structured eDIG observations, 
as those presented in this study.

\section{SUMMARY}
\label{s:sum}

We have observed four edge-on spiral galaxies with different levels of star
formation in H$\alpha$ with the {\it HST}/ACS to study the morphology of 
their extraplanar DIG (eDIG). 
The distribution of the eDIG is very different among the sample galaxies. 
While the eDIG is very filamentary in NGC\,4700 and in NGC\,7090, the 
distribution of the eDIG in NGC\,4634 and NGC\,5775 is remarkably smooth. 
The latter two galaxies have the highest SF rate per unit area in our sample 
and NGC\,5775 is also the most massive galaxy in the sample. 

Based on our current and earlier results, we conclude that the observed
differences are probably a consequence of their different levels of SF activity 
{\it and} their total masses and/or mass densities. 
Whereas the SF rate per unit area controls the presence (e.g., the radial
extent), shape and amount of eDIG in a galaxy's halo, the total mass or mass density
seems to be at least in part responsible for the actual shape of the eDIG. This is 
corroborated by a comparison of the morphology of the lower-mass galaxies to those of 
SF dwarf galaxies such as NGC\,55 or the blue compact dwarf IZw18. The more 
massive galaxies with active SF seem to expell their gaseous outflows more 
globally and diffusively.    

The studied galaxies can be arranged into a {\it sequence} of eDIG diffuseness 
as a function of star formation activity per unit area. More generally speaking, 
there is a morphological DIG sequence among SF galaxies, which evolves from 
galaxies which reveal mostly bubbles and supershells to galaxies which show 
filamentary emission to cases of intense SF which reveal smooth eDIG layers 
(i.e. halos).


\acknowledgments Support for proposal 10416 was provided by NASA through a 
grant from the Space Telescope Science Institute, which is operated by the 
Association of Universities for Research in Astronomy, Inc., under NASA 
contract NAS 5-26555. This project has been supported at Ruhr-University Bochum 
by DLR grant 50 OR 0503. This research has made use of the NASA/IPAC Extragalactic 
Database (NED) which is operated by the Jet Propulsion Laboratory, California Institute 
of Technology, under contract with the National Aeronautics and Space Administration. 
This research has made use of NASA's Astrophysics Data System Bibliographic Services. 
We also made use of the Lyon Extragalactic database (LEDA).

\clearpage


\clearpage



\clearpage

\clearpage

\clearpage

\clearpage

\clearpage

\clearpage

\clearpage

\clearpage

\clearpage

\clearpage



\begin{deluxetable}{lcccccc}
\tabletypesize{\scriptsize}
\tablecaption{Basic Galaxy Sample Properties\label{t:obs}}
\tablewidth{0pt}
\tablehead{
Galaxy & R.A. (J2000.0) & Decl. (J2000.0) & Classification & $cz$ & $D$ & 
$A_V$ \\
 & & & & (km~s$^{-1}$) & (Mpc) & (mag) \\
\colhead{(1)} & \colhead{(2)} & \colhead{(3)} & \colhead{(4)} &
\colhead{(5)} & \colhead{(6)} & \colhead{(7)} \\ 
}
\startdata
\object{NGC\,4634} & 12 42 41.00 & $+$14 17 46.0 & SBcd & 221 & 19.1 & 0.093 \\
\object{NGC\,4700} & 12 49 07.32 & $-$11 24 45.6 & SB(s)c & 1386 & 24.0 & 
0.156 \\
\object{NGC\,5775} & 14 53 57.60 & $+$03 32 40.1 & SBc & 1768 & 26.7 & 0.139 \\
\object{NGC\,7090} & 21 36 28.61 & $-$54 33 26.2 & SBc & 642 & 11.4 & 0.076 \\
\enddata
\tablecomments{Col.~(1): Galaxy name. Col.~(2) and~(3): Right ascension and 
declination from the NASA Extragalactic Database (NED). Units of right 
ascension are hours, minutes and seconds, and units of declination are 
degrees, arcminutes and arcseconds. Col.~(4): Morphological classification 
given by NED. Col.~(5): Systemic velocity corrected for Virgocentric infall 
using the model of \citet{san90}, as taken from LEDA. Col.~(6): Distance, 
taken as listed in \citet{ros00,ros03b}. Col.~(7): Galactic foreground 
extinction \citep{sch98} in the $V$-band for $R_V = 3.1$ from NED.}  
\end{deluxetable}

\clearpage

\begin{deluxetable}{lccccccccc}
\tabletypesize{\scriptsize}
\tablecaption{Journal of ACS Observations\label{t:exp}}
\tablewidth{0pt}
\tablehead{
Galaxy & Data set ident. & Obs. Date & ACS-channel & Filter & 
$\lambda_{\rm{cen}}$ & FWHM & $t_{\rm{int}}$ & Dither & P.A. \\
 & & & & & ({\AA}) & ({\AA}) & (sec) & & ($\degr$) \\
\colhead{(1)} & \colhead{(2)} & \colhead{(3)} & \colhead{(4)} &
\colhead{(5)} & \colhead{(6)} & \colhead{(7)} & \colhead{(8)} & 
\colhead{(9)} & \colhead{(10)} \\
}
\startdata
NGC\,4634 & j95x01 & 2005 Jun 24 & WFC & F625W & 6318 & 1442 & 2308.0 & 
$2\times2$ & 125.596 \\
NGC\,4634 & j95x01 & 2005 Jun 24 & WFC & F658N & 6584 & 78 & 6888.0 & 
$3\times4$ & 125.596 \\
NGC\,4700 & j95x02 & 2005 Jul 30 & WFC & F625W & 6318 & 1442 & 2308.0 & 
$2\times2$ & 115.209 \\
NGC\,4700 & j95x02 & 2005 Jul 30 & WFC & F658N & 6584 & 78 & 6888.0 & 
$3\times4$ & 115.209 \\
NGC\,5775 & j95x03 & 2005 Aug 21 & WFC & F625W & 6318 & 1442 & 2292.0 & 
$2\times2$ & 101.544 \\
NGC\,5775 & j95x03 & 2005 Aug 21 & WFC & F658N & 6584 & 78 & 6848.0 & 
$3\times4$ & 101.544 \\
NGC\,7090 & j95x04 & 2005 Jun 23 & WFC & F625W & 6318 & 1442 & 2508.0 & 
$2\times2$ & $-$72.289 \\
NGC\,7090 & j95x04 & 2005 Jun 23 & WFC & F658N & 6584 & 78 & 7496.0 & 
$3\times4$ & $-$72.289 \\
\enddata
\tablecomments{Col.~(1): Galaxy name. Col.~(2) Data set identifier in the 
{\it HST} Data Archive. Col.~(3): Date of observations. Col.~(4): ACS-aperture 
(WFC: Wide Field Channel). Col.~(5), (6) and (7): Filter properties, namely 
the identifier, central wavelength and full width at half maximum, 
respectively. Col.~(8) Total exposure time, which is the sum of the individual 
dithered images. Col.~(9) Used dither pattern (primary- and sub-pattern). 
Col.~(10): Orientation angle at which the observations were performed.}  
\end{deluxetable}

\begin{deluxetable}{lcccc}
\tabletypesize{\scriptsize}
\tablecaption{Selected Specific Galaxy Properties\label{t:prop}}
\tablewidth{0pt}
\tablehead{
Property & NGC\,4634 & NGC\,4700 & NGC\,5775 & NGC\,7090 \\
}
\startdata
$M_{\rm tot} [\rm{M}_{\sun}]$ & $2.72\times10^{10}$ & $1.91\times10^{10}$ & 
$1.91\times10^{11}$ & $4.52\times10^{10}$ \\
Disk morphology & regular? & irregular & regular & asymmetric \\
eDIG morphology & diffuse & filamentary & diffuse & filamentary \\
Interaction & yes & no & yes & no \\
SF distribution & extended & extended & extended & part disk \\
$SFR_{\rm tot} [\rm{M}_{\sun}\,\rm{yr}^{-1}]$ & 1.08 & 1.00 & 9.86 & 0.58 \\
$S_{60}/S_{100}$ & 0.3720 & 0.5074 & 0.4240 &0.4129 \\
$L_{\rm{FIR}}/D^2_{25} [10^{40}\,\rm{erg\,s^{-1}\,kpc^{-2}}]$  & 11.9 & 5.0 
& 29.6 & 2.4\\
X-ray halo & yes & ... & yes & yes? \\
Radio halo & extended disk & yes & yes & yes \\
$B_{\perp}$ & ... & ... & yes & yes \\
\enddata
\tablecomments{The total galaxy mass $M_{\rm tot}$ in units of solar masses, 
listed in Row~(1) was calculated from \hbox{H\,{\sc i}} measurements using the 
relation $M_i = 3\times10^{4}*A(0)*(\Delta v_0^i)^2$. A(0) is calculated 
either from $D_{25}$ or D[\hbox{H\,{\sc i}}] and has units of kpc, and $\Delta 
v_0^i$ is the inclination corrected full width W[\hbox{H\,{\sc i}}] in units 
of km/s. Row~(2) describes the disk morphology and Row~(3) is the eDIG 
morphology. Row~(4) describes whether the galaxy is interacting or not. 
Row~(5) lists the SF distribution. Row~(6) lists the star formation rate. The 
SFR's are calculated based on FIR data and using the relatation given by 
\citet{ken98}. Row~(7) is the IRAS flux ratio at $60\mu$m and $100\mu$m. These 
are based on the newly calibrated data from \citet{san03}. Row~(8) lists the 
SFR per unit area, expressed as FIR luminosity ($L_{\rm{FIR}}$) divided by the 
diameter of the optical 25th magnitude isophote diameter squared. Row~(9) and 
(10) list whether the galaxy has detections of extended emission in the X-ray 
and radio regime, respectively. The preliminary X-ray analysis of shallow 
XMM-Newton data of NGC\,7090 shows the indication of an X-ray halo. This has 
to be confirmed by deeper observations, which are already scheduled. For 
NGC\,4700 X-ray observations are still pending. No radio halo was detected in 
NGC\,4634, but based on low-resolution data obtained with the 100m Effelsberg 
radio telescope at 2.8\,cm, at least an extended disk is revealed (J. Rossa, 
unpublished). Finally, Row~(11) lists measured magnetic field vectors 
perpendicular to the disk.}  
\end{deluxetable}

\begin{deluxetable}{lccc}
\tabletypesize{\scriptsize}
\tablecaption{Description of the Observed DIG Morphology\label{t:morph}}
\tablewidth{0pt}
\tablehead{
Galaxy & Region & Morphology & Ref. Figure \\
\colhead{(1)} & \colhead{(2)} & \colhead{(3)} & \colhead{(4)} \\
}
\startdata
NGC\,4634 & A & mostly diffuse & Figure~\ref{f:n4634sixpanel}b\\
& & many compact SF regions &  \\
& & few shells SF regions &  \\
 & B & mostly diffuse & Figure~\ref{f:n4634sixpanel}c\\
& & dIrr galaxy to the SW &  \\
 & C & mostly diffuse & Figure~\ref{f:n4634sixpanel}d\\
& & bright shells/supershells interspersed with dust clouds &  \\
 & D & mostly diffuse & Figure~\ref{f:n4634sixpanel}e\\
& & bright SF region (top) &  \\
& & small-scale filaments surrounding bright shells &  \\
& & prominent dust cloud &  \\
 & E & two isolated shells/SF regions & Figure~\ref{f:n4634sixpanel}f\\
& & off-planar feature (to the SW) & \\
\hline \\
NGC\,4700 & A & many superbubbles & Figure~\ref{f:n4700sixpanel}b\\
& & filaments protruding from bright shells &  \\
 & B & superbubbles & Figure~\ref{f:n4700sixpanel}c\\
& & very filamentary &  \\
& & bright supershell (to the SW) &  \\
 & C & spectacular central ``radio lobe'' structure & Figure~\ref{f:n4700sixpanel}d\\
& & filaments protruding from central region &  \\
& & compact SF regions scattered (mostly off-planar) &  \\
 & D & southern extension of central structure & Figure~\ref{f:n4700sixpanel}e\\
& & filamentary + bright shells &  \\
 & E & superbubbles & Figure~\ref{f:n4700sixpanel}f\\
& & few filaments &  \\
\hline \\
NGC\,5775 & A & mostly diffuse & Figure~\ref{f:n5775sixpanel}b\\
& & several bright SF regions of various sizes &  \\
 & B & mostly diffuse & Figure~\ref{f:n5775sixpanel}c\\
& & few bright SF regions of various sizes &  \\
 & C & bright SF regions & Figure~\ref{f:n5775sixpanel}d\\
& & patchy dust regions interspersed &  \\
 & D & faint diffuse emission & Figure~\ref{f:n5775sixpanel}e\\
& & few bright SF regions of various sizes &  \\
 & E & very bright and complex SF region & Figure~\ref{f:n5775sixpanel}f\\
& & diffuse surrounding breaking up into filaments &  \\
\hline \\
NGC\,7090 & A & few bright filamentary SF regions & Figure~\ref{f:n7090sixpanel}f\\
& & many smaller SF regions interspersed &  \\
 & B & complex filamentary SF region around bright star & Figure~\ref{f:n7090sixpanel}e\\
& & few smaller SF regions interspersed &  \\
 & C & very bright SF regions & Figure~\ref{f:n7090sixpanel}d\\
& & copious filaments emanating from SF regions &  \\
& & few compact off-planar SF regions  &  \\
 & D & mainly bubbles and superbubbles & Figure~\ref{f:n7090sixpanel}c\\
& & faint diffuse DIG superimposed &  \\
 & E & few bright SF regions & Figure~\ref{f:n7090sixpanel}b\\
& & bubbles, superbubbles, complex filaments &  \\
\enddata
\tablecomments{Col.~(1) lists the galaxy identifier. Col.~(2) lists the specific 
region under study within each galaxy (labeled by identifiers A,...,E), and with 
the described morphology in Col.~(3) as revealed in the corresponding figures, 
indicated by Col.~(4). The reference figure labeling refers to the following subpanel 
positions within the individual figures: b = top right, c = middle left, d = middle right, 
e = bottom left, f = bottom right.}  
\end{deluxetable}



\end{document}